\documentclass[12pt]{iopart}
\usepackage{graphicx}

\begin{document}
\small 
\title{Functional integral treatment of some quantum 
nondemolition systems} 

\author{Subhashish Banerjee}
\ead{subhashishb@rri.res.in}
\address{Raman Research Institute, Bangalore - 560 080, 
India} 

\author{R Ghosh} 
\ead{rghosh@mail.jnu.ac.in} 
\address{School of Physical Sciences, Jawaharlal Nehru 
University, New Delhi - 110 067, India} 

\begin{abstract}
In the scheme of a quantum nondemolition (QND) measurement, an 
observable is measured without perturbing its evolution. In the 
context of studies of decoherence in quantum computing, we 
examine the `open' quantum system of a two-level atom, or 
equivalently, a spin-1/2 system, in interaction with quantum 
reservoirs of either oscillators or spins, under the QND 
condition of the Hamiltonian of the system commuting with the 
system-reservoir interaction. For completeness, we also examine 
the well-known non-QND spin-Bose problem. For all these 
many-body systems, we use the methods of functional integration 
to work out the propagators. The propagators for the QND 
Hamiltonians are shown to be analogous to the squeezing and 
rotation operators, respectively, for the two kinds of baths 
considered. Squeezing and rotation being both phase space 
area-preserving canonical transformations, this brings out an 
interesting connection between the energy-preserving QND 
Hamiltonians and the homogeneous linear canonical 
transformations. 
\end{abstract} 



\maketitle

\section{Introduction}

In the scheme of a quantum nondemolition (QND) measurement, an 
observable is measured without perturbing its free motion. Such 
a scheme was originally introduced in the context of the 
detection of gravitational waves \cite{caves80}. It was to 
counter the quantum mechanical unpredictability that in general 
would disturb the system being measured. The dynamical 
evolution of a system immediately following a measurement 
limits the class of observables that may be measured repeatedly 
with arbitrary precision, with the influence of the measurement 
apparatus on the system being confined strictly to the 
conjugate observables. Observables having this feature are 
called QND or back-action evasion observables \cite{bo96, vo98, 
zu84}. In addition to its relevance in ultrasensitive 
measurements, a QND scheme provides a way to prepare quantum 
mechanical states which may otherwise be difficult to create, 
such as Fock states with a specific number of particles. One of 
the original proposals for a quantum optical QND scheme was 
that involving the Kerr medium \cite{walls}, which changes its 
refractive index as a function of the number of photons in the 
`signal' pump laser. The advent of experimental methods for 
producing Bose-Einstein condensation (BEC) enables us to make 
progress in the matter-wave analogue of the optical QND 
experiments. In the context of research into BEC, QND schemes 
with atoms are particularly valuable, for instance, in 
engineering entangled states or Schr\"{o}dinger's cat states. A 
state preparation with BEC has recently been performed in the 
form of squeezed state creation in an optical lattice 
\cite{science01}. In a different context, it has been shown 
that the accuracy of atomic interferometry can be improved by 
using QND measurements of the atomic populations at the inputs 
to the interferometer \cite{kbm98}. 

No system of interest, except the entire universe, can be 
thought of as an isolated system -- all subsets of the universe 
are in fact `open' systems, each surrounded by a larger system 
constituting its environment. The theory of open quantum 
systems provides a natural route for reconciliation of 
dissipation and decoherence with the process of quantization. 
In this picture, friction or damping comes about by the 
transfer of energy from the `small' system (the system of 
interest) to the `large' environment. The energy, once 
transferred, disappears into the environment and is not given 
back within any time of physical relevance. Ford, Kac and Mazur 
\cite{fkm65} suggested the first microscopic model describing 
dissipative effects in which the system was assumed to be 
coupled to a reservoir of an infinite number of harmonic 
oscillators. Interest in quantum dissipation, using the 
system-environment approach, was intensified by the works of 
Caldeira and Leggett \cite{cl83}, and Zurek \cite{wz91} among 
others. The path-integral approach, developed by Feynman and 
Vernon \cite{fv63}, was used by Caldeira and Leggett 
\cite{cl83}, and the reduced dynamics of the system of interest 
was followed taking into account the influence of its 
environment, quantified by the influence functional. In the 
model of the fluctuating or ``Brownian" motion of a quantum 
particle studied by Caldeira and Leggett \cite{cl83}, the 
coordinate of the particle was coupled linearly to the harmonic 
oscillator reservoir, and it was also assumed that the system 
and the environment were initially factorized. The treatment of 
the quantum Brownian motion has since been generalized to the 
physically reasonable initial condition of a mixed state of the 
system and its environment by Hakim and Ambegaokar \cite{ha85}, 
Smith and Caldeira \cite{sc87}, Grabert, Schramm and Ingold 
\cite{gsi88}, and by us for the case of a system in a 
Stern-Gerlach potential \cite{sb00}, and also for the quantum 
Brownian motion with nonlinear system-environment couplings 
\cite{sb03-2}. 

An open system Hamiltonian is of the QND type if the 
Hamiltonian $H_S$ of the system commutes with the Hamiltonian 
$H_{SR}$ describing the system-reservoir interaction, i.e., 
$H_{SR}$ is a constant of motion generated by $H_S$. 
Interestingly, such a system may still undergo decoherence or 
dephasing without any dissipation of energy \cite{gkd01,sgc96}. 

In this paper, we study such QND `open system' Hamiltonians of 
particular interest in the context of decoherence in quantum 
computing, and obtain the propagators of the composite systems 
explicitly using path integral methods, for two different 
models of the environment. The aim is to shed some light on the 
problem of QND measurement schemes. Can one draw upon any 
familiar symmetries to connect with the time-evolution 
operation of these QND systems of immense physical importance? 

We take our system to be a two-level atom, or equivalently, a 
spin-1/2 system. We consider two types of environment, 
describable as baths of either oscillators or spins. One cannot 
in general map a spin-bath to an oscillator-bath (or vice 
versa); they constitute distinct 'universality classes' of 
quantum environment \cite{rpp00}. The first case of  
oscillator-bath models (originated by Feynman and Vernon 
\cite{fv63}) describes delocalized environmental modes. For the 
spin-bath, on the other hand, the finite Hilbert space of each 
spin makes it appropriate for describing the low-energy 
dynamics of a set of localized environmental modes. A 
difficulty associated with handling path integrals for spins 
comes from the discrete matrix nature of the spin-Hamiltonians. 
This difficulty is overcome by bosonizing the Hamiltonian by 
representing the spin angular momentum operators in terms of 
boson operators following Schwinger's theory of angular 
momentum \cite{schwin}. 

We then use the Bargmann representation \cite{vb47} for all the 
boson operators. The Schr\"{o}dinger representation of quantum 
states diagonalizes the position operator, expressing pure 
states as wave functions, whereas the Bargmann representation 
diagonalizes the creation operator $b^{\dagger}$, and expresses 
each state vector $|\psi \rangle$ in the Hilbert state ${\cal 
H}$ as an entire analytic function $f(\alpha)$ of a complex 
variable $\alpha$. The association $|\psi \rangle 
\longrightarrow f(\alpha)$ can be written conveniently in terms 
of the normalized coherent states $|\alpha \rangle$ which are 
the right eigenstates of the annihilation operator $b$: 
\begin{eqnarray}
b|\alpha \rangle & = & \alpha |\alpha \rangle , \nonumber \\ 
\langle \alpha '|\alpha \rangle & = & \exp \left( -\frac{1}{2}| 
\alpha '|^2 - \frac{1}{2}|\alpha |^2 + \alpha '^* \alpha 
\right) , \nonumber 
\end{eqnarray}
giving
\[ f(\alpha ) = e^{-|\alpha |^2/2} ~\langle \alpha ^* |\psi 
\rangle . \] 
We obtain the explicit propagators for these many-body systems 
from those of the expanded bosonized forms by appropriate 
projection. 

The propagators for the QND Hamiltonians with an oscillator 
bath and a spin bath are shown to be analogous to the squeezing 
and rotation operators, respectively, which are both phase 
space area-preserving canonical transformations. This suggests 
an interesting connection between the energy-preserving QND 
Hamiltonians and the homogeneous linear canonical 
transformations, which would need further systematic probing. 

The plan of the paper is as follows. In section 2 we take up 
the case of a QND-type of open system Hamiltonian where the 
bath is a bosonic one of harmonic oscillators. In section 2.1 
we consider a case, which is a variant of the previous one, 
wherein we include an external mode in resonance with the 
atomic transition and obtain its propagator. In section 2.2 we 
discuss the non-QND variant of the Hamiltonian which usually 
occurs in the literature in discussions of the spin-Bose 
problem \cite{papa86, lc87}. In section 3 we treat the case 
of a QND-type of open system Hamiltonian where the bath is 
composed of two-level systems or spins. The structure of the 
propagators in the two cases of the oscillator and spin baths 
is discussed in section 4, and in section 5 we present our 
conclusions. 

\section{Bath of harmonic oscillators} 

We first take the case where the system is a two-level atom 
interacting with a bosonic bath of harmonic oscillators with a 
QND type of coupling. Such a model has been studied 
\cite{unruh95, ps96, dd95} in the context of the influence of 
decoherence in quantum computation. The total system evolves 
under the Hamiltonian, 
\begin{eqnarray}
H_1 & = & H_S + H_R + H_{SR} \nonumber\\ & = & {\hbar \omega 
\over 2} \sigma_z + \sum\limits^M_{k=1} \hbar \omega_k 
b^{\dagger}_k b_k + \left( {\hbar \omega \over 2} \right) 
\sum\limits^M_{k=1} g_k (b_k + b^{\dagger}_k) \sigma_z.   
\label{h1} 
\end{eqnarray}
Here $H_S, H_R$ and $H_{SR}$ stand for the Hamiltonians of the 
system, reservoir, and system-reservoir interaction, 
respectively. We have made use of the equivalence of a 
two-level atom and a spin-1/2 system, $\sigma_x, \sigma_z$ 
denote the standard Pauli spin matrices and are related to the 
spin-flipping (or atomic raising and lowering) operators $S_+$ 
and $S_-$: $\sigma_x = S_+ + S_-$, $\sigma_z = 2 S_+ S_- - 1$. 
In (\ref{h1}) $b^{\dagger}_k, b_k$ denote the Bose creation and 
annihilation operators for the $M$ oscillators of frequency 
$\omega_k$ representing the reservoir, $g_k$ stands for the 
coupling constant (assumed real) for the interaction of the 
field with the spin. Since $[H_S, H_{SR}]=0$, the Hamiltonian 
(\ref{h1}) is of QND type. 

The explicit propagator $\exp (-{i H t \over \hbar})$ for the 
Hamiltonian (\ref{h1}) is obtained by using functional 
integration and bosonization \cite{papa86,sb03-1}. In order to 
express the spin angular momentum operators in terms of boson 
operators, we employ Schwinger's theory of angular momentum 
\cite{schwin} by which any angular momentum can be represented 
in terms of a pair of boson operators with the usual 
commutation rules. The spin operators $\sigma_z$ and $\sigma_x$ 
can be written in terms of the boson operators $a_{\beta}$, 
$a_{\beta}^{\dagger}$ and $a_{\gamma}$, $a_{\gamma}^{\dagger}$ 
as 
\begin{eqnarray}
\sigma_z & = & a_{\gamma}^{\dagger} a_{\gamma} -  
a_{\beta}^{\dagger} a_{\beta}, \nonumber \\ \sigma_x & = & 
a_{\gamma}^{\dagger} a_{\beta} + a_{\beta}^{\dagger} a_{\gamma} 
. \nonumber 
\end{eqnarray}
In the Bargmann representation \cite{vb47} the actions of $b$ 
and $b^{\dagger}$ are 
\begin{eqnarray}
b^{\dagger} f(\alpha) & = & \alpha^* f(\alpha), \nonumber\\ b 
f(\alpha) & = & {df(\alpha) \over d\alpha^*},   
\end{eqnarray}
where $|\alpha \rangle$ is the normalized coherent state. The 
spin operator becomes
\begin{equation}
\sigma_z \longrightarrow \left(  \gamma^* {\partial \over 
\partial \gamma^*} - \beta^* {\partial \over \partial \beta^*} 
\right).   
\end{equation}
Here the variable $\beta^*$ is associated with the spin-down 
state and the variable $\gamma^*$ with the spin-up state. 

The bosonized form of the Hamiltonian (1) is
\begin{eqnarray}
H_{B_{1}} & = & {\hbar \omega \over 2} \left( \gamma^* 
{\partial \over \partial \gamma^*} - \beta^* {\partial \over 
\partial \beta^*} \right) + \sum\limits^M_{k=1} \hbar \omega_k 
\alpha^*_k {\partial \over \partial \alpha^*_k} \nonumber\\ & & 
+ {\hbar \omega \over 2} \sum\limits^M_{k=1} g_k \left( 
\alpha^*_k + {\partial \over \partial \alpha^*_k} \right) 
\left( \gamma^* {\partial \over \partial \gamma^*} - \beta^* 
{\partial \over \partial \beta^*} \right). 
\end{eqnarray}
Here $\alpha^*_k$, ${\partial \over \partial \alpha^*_k}$ are 
the Bargmann representations for $b^{\dagger}_k$ and $b_k$, 
respectively. A particular solution of the Schr$\ddot{o}$dinger 
equation for the bosonized Hamiltonian (4) is 
\begin{equation}
U_1 = U_{00} \beta^* \beta' + U_{01} \beta^* \gamma' + U_{10} 
\gamma^* \beta' + U_{11} \gamma^* \gamma', 
\end{equation}
where the amplitude $U_{ij}$ are functions of time as well as 
the coherent state variables associated with the boson 
oscillators, with the initial condition 
\begin{equation}
U_{ij} (t=0) = \exp \left\{ \sum\limits^M_{k=1} \alpha^*_k 
\alpha'_k \right\} \delta_{ij} ~~~~~(i,j=0,1). 
\end{equation}
The initial state for the expanded propagator associated with 
the bosonized Hamiltonian (5) is 
\begin{equation}
U(t=0) = \exp \left\{ \sum\limits^M_{k=1} \alpha^*_k \alpha'_k 
\right\} \exp \left\{ \beta^* \beta' + \gamma^* \gamma' 
\right\}. 
\end{equation}
If the Hamiltonian is in the normal form given by $H(\alpha^*, 
{\partial \over \partial \alpha^*}, t)$, the associated 
propagator is given as a path integral over coherent state 
variables as \cite{klauder} 
\begin{equation}
U(\alpha^*,t;\alpha',0) = \int {\bf D} \{ \alpha \} \exp 
\left\{ \sum\limits_{0\leq \tau < t} \alpha^* (\tau +) \alpha 
(\tau) - {i \over \hbar} \int\limits^t_0 d\tau H \left( 
\alpha^* (\tau +), \alpha (\tau), \tau \right) \right\}. 
\end{equation}
Here $\sum\limits_{0\leq \tau <t} \alpha^* (\tau +) \alpha 
(\tau)$ stands for $\sum\limits^{N-1}_{j=0} \alpha^* 
(\tau_{j+1}) \alpha (\tau_j)$ in the subdivision of the 
internal $[0,t]$, i.e., where $\tau$ stands for $\tau_j$, $\tau 
+$ stands for the next point $\tau_{j+1}$ in the subdivision. 
Also, in the subdivision scheme, $$ \int\limits^t_0 d\tau 
H\left( \alpha^* (\tau +), \alpha (\tau), \tau \right) = 
\sum\limits^{N-1}_{j=0} H\left( \alpha^* (\tau_{j+1}), \alpha 
(\tau_j), \tau_j \right) \Delta \tau_j. $$ Here the path 
differential in (8) is 
\begin{equation}
{\bf D}^2 \{ \alpha \} = \prod_{0<\tau <t} D^2 \alpha (\tau), 
\end{equation}
where the weighted differential is
\begin{equation}
D^2 \alpha (\tau) = {1 \over \pi} \exp \left( -|\alpha 
(\tau)|^2 \right) d^2 \alpha (\tau). 
\end{equation}
Using (8), the propagator for the bosonized Hamiltonian (4) is 
\begin{eqnarray}
u_1 (\mbox{\boldmath $\alpha^*$}, \beta^*, \gamma^*, t; 
\mbox{\boldmath $\alpha'$}, \beta', \gamma', 0) & = & \int {\bf 
D}^2 \{ \mbox{\boldmath $\alpha$} \} {\bf D}^2 \{\beta\} {\bf 
D}^2 \{\gamma\} \nonumber \\ & & \times \exp \Bigg\{ 
\sum\limits_{0\leq \tau <t} \Bigg[  \sum\limits^M_{k=1} 
\alpha^*_k (\tau+) \alpha_k (\tau) \nonumber \\ & & + \beta^* 
(\tau+) \beta (\tau) + \gamma^* (\tau+) \gamma (\tau) \Bigg] 
\nonumber \\ & & - i\sum\limits^M_{k=1} \int\limits^t_0 d\tau 
\omega_k \alpha^*_k (\tau+) \alpha_k (\tau) \nonumber \\ & & - 
i {\omega \over 2} \int\limits^t_0 d\tau \Bigg[ \gamma^*(\tau+) 
\gamma (\tau) - \beta^* (\tau+) \beta(\tau) \Bigg] \nonumber\\ 
& & - i {\omega \over 2} \sum\limits^M_{k=1} \int\limits^t_0 
d\tau g_k \Bigg[  \alpha^*_k (\tau+) + \alpha_k (\tau) \Bigg] 
\Bigg[ \gamma^* (\tau+) \gamma(\tau) \nonumber \\ & & - 
\beta^*(\tau+) \beta(\tau) \Bigg] \Bigg\} . 
\end{eqnarray}
In Eq. (11) {\boldmath $\alpha$} is a vector with components 
$\{ \alpha_k\}$, and ${\bf D}^2\{ \mbox{\boldmath $\alpha$}\} = 
\prod^M_{k=1} {\bf D}^2 \{ \alpha_k \}$. 

Now we introduce a complex auxiliary field $f(\tau)$ to 
decouple the interaction term in (11) as 
\begin{eqnarray}
& & \exp \left( -{i\omega \over 2} \sum\limits^M_{k=1} 
\int\limits^t_0 d\tau g_k \left[ \alpha^*_k (\tau +) + \alpha_k 
(\tau) \right] \left[ \gamma^* (\tau +) \gamma (\tau) - \beta^* 
(\tau +) \beta (\tau) \right] \right) \nonumber\\ & & = \int 
{\bf D}^2 \{ f\} \exp \left[ -i \sum\limits^M_{k=1} 
\int\limits^t_0 d\tau f^* (\tau) g_k \left( \alpha^*_k (\tau +) 
+ \alpha_k (\tau) \right) \right] \nonumber \\ & & \times \exp 
\left[ \int\limits^t_0 d\tau f(\tau) {\omega \over 2} \left( 
\gamma^* (\tau +) \gamma (\tau) - \beta^* (\tau +) \beta (\tau) 
\right) \right]. 
\end{eqnarray}
Here we have used the $\delta$-functional identify, 
\cite{papa86} 
\begin{equation}
\int {\bf D}^2 \{ x \} P[x^*(t)] \exp \left\{ \int\limits^t_0 
d\tau y(\tau)x(\tau) \right\} = P[y(t)], 
\end{equation}
where ${\bf D}^2\{ x\}$ is the functional differential
\begin{equation}
{\bf D}^2 \{ x\} = \exp \left( -\int\limits^t_0 d\tau 
|x(\tau)|^2 \right) \prod_{0\leq \tau <t} \left( {d\tau \over 
\pi} \right) d^2 x(\tau), 
\end{equation}
and ${\bf P}[x^*(t)]$ is an explicit functional of $x^*$ only. 
Using (12), the bosonized propagator (11) can be written as 
\begin{eqnarray}
u_1 (\mbox {\boldmath $ \alpha^* $}, \beta^*, \gamma^*, t; 
\mbox{\boldmath $\alpha'$}, \beta', \gamma', 0) & = & \int {\bf 
D}^2 \{f\} G_1 (\mbox{\boldmath $\alpha^*$}, t; \mbox{\boldmath 
$\alpha'$}, 0; [f^*]) \nonumber \\ & & \times N_1 \left( 
\beta^*, \gamma^*, t; \beta', \gamma', 0; [f] \right). 
\end{eqnarray}
Here $G_1$ stands for the propagator for
\begin{equation}
H_{G_{1}} = \hbar \sum\limits^M_{k=1} \left[ \omega_k 
\alpha^*_k {\partial \over \partial \alpha^*_k} + f^*(t) g_k 
\alpha^*_k + f^*(t) g_k \alpha_k \right], 
\end{equation}
$N_1$ is the propagator for 
\begin{equation}
H_{N_{1}} = {\hbar \omega \over 2}\left( \gamma^* {\partial 
\over \partial \gamma^*} - \beta^* {\partial \over \partial 
\beta^*} \right) + {i\hbar \omega \over 2} f(t) \left( \gamma^* 
{\partial \over \partial \gamma^*} - \beta^* {\partial \over 
\partial \beta^*} \right). 
\end{equation}
These obey the Schr\"{o}dinger equations $i\hbar {\partial 
\over \partial t} G_1=H_{G_{1}} G_1, i\hbar {\partial \over 
\partial t} N_1=H_{N1} N_1$ with the initial conditions 
\begin{eqnarray}
G_1 (t=0) & = & \exp \left\{ \sum\limits^M_{k=1} \alpha^*_k 
\alpha'_k \right\}, \nonumber\\ N_1 (t=0) & = & \exp \left\{ 
\beta^* \beta' + \gamma^* \gamma' \right\}. 
\end{eqnarray}
The propagator $G_1$ is given by
\begin{eqnarray}
G_1 & = & \exp \Bigg\{ \sum\limits^M_{k=1} \alpha^*_k \alpha'_k 
e^{-i\omega_kt} - \sum\limits^M_{k=1} \Bigg[ i \alpha^*_k g_k 
\int\limits^t_0 d\tau f^* (\tau) e^{-i\omega_k(t-\tau)} 
\nonumber\\ & & + i\alpha'_k g_k \int\limits^t_0 d\tau e^{-
i\omega_k\tau} f^* (\tau) \nonumber \\ & & + g^2_k 
\int\limits^t_0 d\tau \int\limits^{\tau}_0 d\tau' e^{-i\omega_k 
(\tau-\tau')} f^* (\tau) f^* (\tau') \Bigg] \Bigg\}. 
\end{eqnarray}
The propagator $N_1$ is given by
\begin{eqnarray}
N_1 & = & \exp \left\{ Q_{00} \beta^* \beta' + Q_{01} \beta^* 
\gamma' + Q_{10} \gamma^* \beta' + Q_{11} \gamma^* \gamma' 
\right\} \nonumber \\ & = & \sum\limits^{\infty}_{l=0} {1 \over 
l!} \left[ (\beta^*,\gamma^*) Q \pmatrix{\beta' \cr \gamma'} 
\right]^l. 
\end{eqnarray}
Here $Q(t)$ is given by
\begin{equation}
Q(t) = \exp \left( {i\omega \over 2} \sigma_z t - {\omega \over 
2} \sigma_z \int\limits^t_0 d\tau f(\tau) \right) ,
\end{equation}
with $Q_{ij}(0) = \delta_{ij}, Q(0)=I$.

Thus the propagator for the bosonized Hamiltonian (4) as given 
by (15) becomes 
\begin{equation}
u_1 = \sum\limits^{\infty}_{l=0} \int {\bf D}^2 \{ f\} G_1 {1 
\over l!} \left[ (\beta^*, \gamma^*) Q \pmatrix{\beta' \cr 
\gamma'} \right]^l. 
\end{equation}
The propagator for the Hamiltonian (1) is obtained from (22) by 
taking the $l=1$ term in the above equation. By making use of 
the $\delta$-functional identity (13) the amplitudes of the 
propagator for the Hamiltonian (1) are obtained in matrix form 
as 
\begin{eqnarray}
u_1 = \pmatrix{U_{00} & U_{01} \cr U_{10} & U_{11}} & = & \exp 
\left\{ \sum\limits^M_{k=1} \alpha^*_k \alpha'_k e^{-
i\omega_kt} \right\} \nonumber\\ & & \times e^A \pmatrix{e^B & 
0 \cr 0 & e^{-B}}, 
\end{eqnarray}
where
\begin{equation}
A = i \left( {\omega \over 2} \right)^2 \sum\limits^M_{k=1} 
{g^2_k \over \omega_k} t - \left( {\omega \over 2} \right)^2 
\sum\limits^M_{k=1} {g^2_k \over \omega^2_k} \left( 1-e^{-
i\omega_kt} \right), 
\end{equation}
\begin{equation}
B = \sum\limits^M_{k=1} \phi_k \left( \alpha^*_k + \alpha'_k 
\right) + i {\omega \over 2} t, 
\end{equation}
\begin{equation}
\phi_k = {\omega \over 2} {g_k \over \omega_k} \left( 1-e^{-
i\omega_kt} \right). 
\end{equation}

Here we associate the values $\alpha^*$ with time $t$ and 
$\alpha'$ with time $t=0$ as is also evident from (8). The 
simple form of the last term on the right-hand side of (23) 
reveals the QND nature of the system-reservoir coupling. Since 
we are considering the unitary dynamics of the complete 
Hamiltonian (\ref{h1}) there is no decoherence, and the 
propagator (23) does not have any off-diagonal terms. In a 
treatment of the system alone, i.e., an open system analysis of 
Eq. (\ref{h1}) after the tracing over the reservoir degrees of 
freedom, it has been shown \cite{ps96} that the population, 
i.e., the diagonal elements of the reduced density matrix of 
the system remain constant in time while the off-diagonal 
elements that are a signature of the quantum coherences decay 
due to decoherence, as expected. 

Note that though the commonly used coordinate-coupling model 
describing a free particle in a bosonic bath, explicitly solved 
by Hakim and Ambegaokar \cite{ha85}, with 
\begin{equation}
H = {P^2 \over 2} + {1 \over 2} \sum\limits^M_{j=1} \left( 
p^2_j + \omega^2_j (q_j - Q)^2 \right) ,  \label{h14} 
\end{equation} 
is seemingly not of the QND type, it can be shown to be 
unitarily equivalent to a Hamiltonian of the QND type as 
follows: 
\begin{eqnarray}
U_2 U_1 H U^{\dagger}_1 U^{\dagger}_2 & = & {P^2 \over 2} + P 
\sum\limits^M_{j=1} \omega_j q_i + {1 \over 2} 
\sum\limits^M_{j=1} \left( p^2_j + \omega^2_j q^2_j \right) 
\nonumber\\ & & + {1 \over 2} \left( \sum\limits^M_{j=1} 
\omega_j q_j \right)^2 , \label{ha1} 
\end{eqnarray}
where $U_1$ and $U_2$ are the unitary operators
\begin{equation}
U_1 = \exp \left[ {i\pi \over 2\hbar} \sum\limits^M_{j=1} 
\left( {p^2_j \over 2\omega_j} + {1 \over 2} \omega_j q^2_j 
\right) \right], 
\end{equation}
\begin{equation}
U_2 = \exp \left[ {-i \over \hbar} Q \sum\limits^M_{j=1} 
\omega_j q_j \right]. 
\end{equation}
The above Hamiltonian (\ref{ha1}) is of the QND type with 
$[H_S, H_{SR}]=[P^2/2, P \sum\limits^M_{j=1} \omega_j q_j]=0$. 
It is commonly known as the velocity-coupling model 
\cite{flc88}. 

\subsection{An external mode in resonance with the atomic 
transition} 

In this subsection we consider a Hamiltonian which is a variant 
of the one in (1): 
\begin{eqnarray}
H_2 & = & {\hbar \omega \over 2} \sigma_z + \hbar \Omega 
a^{\dagger}a - {\hbar \Omega \over 2} \sigma_z \nonumber \\ & & 
+ \sum\limits^M_{k=1} \hbar \omega_k b^{\dagger}_k b_k + {\hbar 
\omega \over 2} \sum\limits^M_{k=1} g_k (b_k + b^{\dagger}_k) 
\sigma_z. 
\end{eqnarray}
Here
\begin{equation}
\Omega = 2 \vec{\epsilon}.\vec{d}^*,
\end{equation}
where $\vec{d}$ is the dipole transition matrix element and 
$\vec{\epsilon}$ comes from the field strength of the external 
driving mode $\vec{E}_L(t)$ such that 
\begin{equation}
\vec{E}_L(t) = \vec{\epsilon} e^{-i\omega t} + \vec{\epsilon}^* 
e^{i\omega t}. 
\end{equation}
Here we have used the form $-{\Omega \over 2} \sigma_z$, 
associated with the external mode, instead of the usual form $-
{\Omega \over 2} \sigma_x$ and (31) is of a QND type. 
Proceeding as in section 2 and introducing the symbol $\nu^*$ 
for the Bargmann representation of the external mode 
$a^{\dagger}$ we have the amplitudes of the propagator for (31) 
in matrix form as 
\begin{eqnarray}
u_2 = \pmatrix{U_{00} & U_{01} \cr U_{10} & U_{11}} & = &  \exp 
\left\{ \sum\limits^M_{k=1} \alpha^*_k \alpha'_k e^{-i\omega_k 
t} \right\} \nonumber\\ & & \times \exp \left\{ \nu^* \nu' e^{-
i\Omega t} \right\} e^A \pmatrix{e^{B_{2}} & 0 \cr 0 & e^{-
B_{2}}}, 
\end{eqnarray}
where $A$ is as in Eq. (24),
\begin{equation}
B_2 = \sum\limits^M_{k=1} \phi_k \left( \alpha^*_k + \alpha'_k 
\right) + i \left( {\omega - \Omega \over 2} \right) t, 
\end{equation}
and $\phi_k$ is as in Eq. (26).

\subsection{Non-QND spin-Bose problem}

In this subsection we consider a Hamiltonian that is a variant 
of the spin-Bose problem \cite{papa86, sb03-1, lc87}. This 
addresses a number of problems of importance such as the 
interaction of the electromagnetic field modes with a two-level 
atom \cite{cm78, rn82}.  Another variant of the spin-Bose 
problem has been used for treating problems of phase 
transitions \cite{be70, cl84} and also to the tunnelling 
through a barrier in a potential well \cite{uw93}. Our 
Hamiltonian is 
\begin{eqnarray}
H_3 & = & {\hbar \omega \over 2} \sigma_z + \sum\limits^M_{k=1} 
\hbar \omega_k b^{\dagger}_k b_k \nonumber\\ & & + {\hbar 
\omega \over 2} \sum\limits^M_{k=1} g_k (b_k + b^{\dagger}_k) 
\sigma_x. \label{p1} 
\end{eqnarray}
This could describe, for example, the interaction of $M$ modes 
of the electromagnetic field with a two-level atom via a dipole 
interaction. This has a form similar to Eq. (1) except that 
here the system-environment coupling is via $\sigma_x$ rather 
than $\sigma_z$. This makes the Hamiltonian (\ref{p1}) a 
non-QND variant of the Hamiltonian (1). We proceed as in 
Section II with $H_{N_{1}}$ (17) now given by 
\begin{equation}
H_{N_{1}} = {\hbar \omega \over 2} \left( \gamma^* {\partial 
\over \partial \gamma^*} - \beta^* {\partial \over \partial 
\beta^*} \right) + i {\hbar \omega \over 2} f(t) \left( 
\gamma^* {\partial \over \partial \beta^*} + \beta^* {\partial 
\over \partial \gamma^*} \right). \label{p2} 
\end{equation}
The propagator for $H_{N_{1}}$ (\ref{p2}) has the same form as 
$N_1$ (20) but with $Q$ now satisfying the equation 
\begin{equation}
{\partial \over \partial t} Q = {i\omega \over 2} \sigma_z Q + 
{\omega \over 2} f(t) \sigma_x Q, 
\end{equation}
with $Q_{ij}(0)=\delta_{ij}, Q(0)=I$. This is solved 
recursively to yield the series solution 
\begin{eqnarray}
Q(t) & = & \sum\limits^{\infty}_{n=0} Q^{(n)} (t), \nonumber \\ 
Q^{(n)} (t) & = & \left( {i\omega \over 2} \sigma_z \right)^n 
\int\limits^t_0 d\tau_n \int\limits^{\tau_n}_0 d\tau_{n-1}... 
\int\limits^{\tau_{2}}_0 d\tau_1 \nonumber\\ & & \times \exp 
\left[ {\omega \over 2} \sigma_x \left( 
\int\limits^{\tau_{1}}_0 - \int\limits^{\tau_{2}}_{\tau_{1}} + 
... + (-1)^n \int\limits^t_{\tau_{n}} \right) d\tau f(\tau) 
\right]. \label{p3} 
\end{eqnarray}
Using Eq. (\ref{p3}) and proceeding as before, we obtain the 
amplitudes of the propagator for the Hamiltonian (\ref{p1}) in 
matrix form as 
\begin{eqnarray}
u_3 & = & \pmatrix{U_{00} & U_{01} \cr U_{10} & U_{11}} 
\nonumber\\ & = & \exp \left\{ \sum\limits^M_{k=1} \alpha^*_k 
\alpha'_k e^{-i\omega_kt} \right\} \nonumber\\ & & \times 
\sum\limits^{\infty}_{n=0} \left( {i\omega \over 2} \right)^n 
\int\limits^t_0 d\tau_n \int\limits^{\tau_{n}}_0 d\tau_{n-1} 
... \int\limits^{\tau_{2}}_0 d\tau_1 \exp \left\{ \kappa^{(n)} 
\right\} \nonumber\\ & & \times \pmatrix{\cosh \left( 
\chi^{(n)} \right) & \sinh \left( \chi^{(n)} \right) \cr (-1)^n 
\sinh \left( \chi^{(n)} \right) & (-1)^n \cosh \left( 
\chi^{(n)} \right)}, 
\end{eqnarray}
where  
\begin{eqnarray}
\kappa^{(n)} & = & - \left( {\omega \over 2} \right)^2 
\sum\limits^M_{k=1} {g^2_k \over \omega^2_k} \Bigg[ (2n+1) - 
i\omega_kt + (-1)^{n+1} e^{-i\omega_kt} \nonumber \\ & & - 2 
\sum\limits^n_{l=1} (-1)^{l+1} e^{-i\omega_k \tau_l} + 2(-1)^n 
\sum\limits^n_{l=1} (-1)^{l+1} e^{-i\omega_k(t-\tau_l)} 
\nonumber \\ & & + 4 \sum\limits^n_{p=2} \sum\limits^{p-1}_{q=1} 
(-1)^{p+q} e^{-i\omega_k (\tau_p-\tau_q)} \Bigg], 
\end{eqnarray}
and
\begin{eqnarray}
\chi^{(n)} & = & - {\omega \over 2} \sum\limits^M_{k=1} {g_k 
\over \omega_k} \Bigg[ \left( \alpha'_k + (-1)^n \alpha^*_k 
\right) \left( 1+(-1)^{n+1} e^{-i\omega_kt} \right) \nonumber 
\\ & & + 2\alpha^*_k \sum\limits^n_{l=1} (-1)^{l+1} e^{-
i\omega_k(t-\tau_l)} - 2\alpha'_k \sum\limits^n_{l=1} (-
1)^{l+1} e^{-i\omega_k\tau_l} \Bigg]. 
\end{eqnarray}
This agrees with the results obtained in \cite{papa86, sb03-1}. 
The matrix on the right-hand side of Eq. (40) contains diagonal 
as well as off-diagonal terms in contrast to the matrix on the 
right-hand side of Eq. (23) in which only diagonal elements are 
present. This is due to the non-QND nature of the system-bath 
interaction of the Hamiltonian described by Eq. (\ref{p1}) 
whose propagator is given by Eq. (40), whereas Eq. (23) is the 
propagator of the Hamiltonian given by Eq. (1) where the 
system-bath interaction is of the QND type. The simpler form of 
the structure of the propagator (23) compared to the non-QND 
propagator (40) reflects on the simplification in the dynamics 
due to the QND nature of the coupling.     

\section{Bath of spins}

Now we consider the case where the reservoir is composed of 
spin-half or two-level systems, as has been dealt with by Shao 
and collaborators in the context of QND systems \cite{sgc96} 
and also quantum computation \cite{sh97}, and for a nanomagnet 
coupled to nuclear and paramagnetic spins \cite{rpp00}. The 
total Hamiltonian is taken as 
\begin{eqnarray}
H_4 & = & H_S + H_R + H_{SR} \nonumber\\ & = & {\hbar \omega 
\over 2} S_z + \sum\limits^M_{k=1} \hbar \omega_k \sigma_{zk} + 
{\hbar \omega \over 2} \sum\limits^M_{k=1} c_k \sigma_{xk}S_z . 
\label{h10} 
\end{eqnarray}
Here we use $S_z$ for the system and $\sigma_{zk}, \sigma_{xk}$ 
for the bath. Since $[H_S, H_{SR}]=0$, we have a QND 
Hamiltonian. In the Bargmann representation, we associate the 
variable $\beta^*$ with the spin-down state and the variable 
$\gamma^*$ with the spin-up state for the bath variables, and 
we have 
\begin{eqnarray}
\sigma_z & \longrightarrow & \gamma^* {\partial \over \partial 
\gamma^*} - \beta^* {\partial \over \partial \beta^*}, 
\nonumber \\ \sigma_x & \longrightarrow & \gamma^* {\partial 
\over \partial \beta^*} + \beta^* {\partial \over \partial 
\gamma^*}. 
\end{eqnarray}
Similarly, the bosonization of the system variable gives 
\begin{equation}
S_z \longrightarrow \xi^* {\partial \over \partial \xi^*} - 
\theta^* {\partial \over \partial \theta^*} , 
\end{equation}
where the variable $\theta^*$ is associated with the spin-down 
state and the variable $\xi^*$ with the spin-up state. The 
bosonized form of the Hamiltonian (43) is given by 
\begin{eqnarray}
H_{B_{4}} & = & {\hbar \omega \over 2} \left( \xi^* {\partial 
\over \partial \xi^*} - \theta^* {\partial \over \partial 
\theta^*} \right) + \sum\limits^M_{k=1} \hbar \omega_k \left( 
\gamma^*_k {\partial \over \partial \gamma^*_k} - \beta^*_k 
{\partial \over \partial \beta^*_k} \right) \nonumber \\ & & + 
{\hbar \omega \over 2} \sum\limits^M_{k=1} c_k \left( 
\gamma^*_k {\partial \over \partial \beta^*_k} + \beta^*_k 
{\partial \over \partial \gamma^*_k} \right) \left( \xi^* 
{\partial \over \partial \xi^*} - \theta^* {\partial \over 
\partial \theta^*} \right). 
\end{eqnarray}
A particular solution of the Schr\"{o}dinger equation for the 
bosonized Hamiltonian (46) is obtained by attaching amplitudes 
to the polynomial parts in the products 
\begin{equation}
U_4 = (\theta^* + \xi^*) (\theta' + \xi') \prod^M_{k=1} \left( 
\beta^*_k + \gamma^*_k \right) \left( \beta'_k + \gamma'_k 
\right). 
\end{equation}
The initial state for the expanded propagator associated with 
the bosonized Hamiltonian (46) is 
\begin{equation}
U (t=0) = \exp \left\{ \theta^* \theta' + \xi^* \xi' \right\} 
\prod^M_{k=1} \exp \left\{ \beta^*_k \beta'_k + \gamma^*_k 
\gamma'_k \right\}. 
\end{equation}
Using (8), the propagator for the bosonized Hamiltonian (46) is 
\begin{eqnarray}
\fl u_4 (\theta^*, \xi^*, \mbox{\boldmath $\beta^*$}, 
\mbox{\boldmath $\gamma^*$}, t; \theta', \xi', \mbox{\boldmath 
$\beta'$}, \mbox{\boldmath $\gamma'$}, 0) & = & \prod^M_{k=1} 
\int {\bf D}^2 \{\theta\} {\bf D}^2 \{\xi \} {\bf D}^2 
\{\beta_k \} {\bf D}^2 \{\gamma_k \} \nonumber \\ & & \times 
\exp \Bigg\{ \sum\limits_{0\leq \tau <t} \Bigg[ \theta^* 
(\tau+) \theta (\tau) + \xi^* (\tau +) \xi (\tau) \nonumber \\ 
& & + \beta^*_k (\tau+) \beta_k (\tau) + \gamma^*_k (\tau+) 
\gamma_k (\tau) \Bigg] \nonumber \\ & & - i {\omega \over 2} 
\int\limits^t_0 d\tau \Bigg[ \xi^*(\tau+) \xi (\tau) - \theta^* 
(\tau+) \theta(\tau) \Bigg] \nonumber \\ & & - i 
\int\limits^t_0 d\tau \omega_k \Bigg[ \gamma^*_k(\tau+) 
\gamma_k (\tau) - \beta^*_k (\tau+) \beta_k(\tau) \Bigg] 
\nonumber \\ & & - i {\omega \over 2} \int\limits^t_0 d\tau c_k 
\Bigg[ \gamma^*_k(\tau+) \beta_k (\tau) + \beta^*_k(\tau+) 
\gamma_k (\tau) \Bigg] \nonumber \\ & & \times \Bigg[ \xi^*(\tau+) 
\xi (\tau) - \theta^* (\tau+) \theta(\tau) \Bigg] \Bigg\}. 
\end{eqnarray}
On the left-hand side of Eq. (49), {\boldmath $\beta^*$}, 
{\boldmath $\gamma^*$} are vectors with components $\{ 
\beta_k\}$ and $\{ \gamma_k\}$, respectively. Now we introduce 
a complex auxiliary field $f(\tau)$ to decouple the interaction 
term in (49) as 
\begin{eqnarray}
& & \exp \left( -i {\omega \over 2} \int\limits^t_0 d\tau c_k 
\left[ \gamma^*_k (\tau +) \beta_k (\tau)  + \beta^*_k (\tau +) 
\gamma_k (\tau) \right] \left[ \xi^* (\tau +) \xi (\tau) - 
\theta^* (\tau +) \theta (\tau) \right] \right) \nonumber \\ & 
& = \int D^2 \{ f\} \exp \left[ -i \int\limits^t_0 d\tau f^* 
(\tau) c_k \left( \gamma^*_k (\tau +) \beta_k (\tau) + 
\beta^*_k (\tau +) \gamma_k (\tau) \right) \right] \nonumber \\ 
& & \times \exp \left[ \int\limits^t_0 d\tau f(\tau) {\omega 
\over 2} \left( \xi^* (\tau +) \xi (\tau) - \theta^* (\tau +) 
\theta (\tau) \right) \right]. 
\end{eqnarray}
Using (50) in (49) the propagator for the bosonized Hamiltonian 
(46) becomes 
\begin{eqnarray}
\fl u_4 \left( 
\theta^*,\xi^*, \mbox{\boldmath $\beta^*$}, 
\mbox{\boldmath $\gamma^*$},t;\theta',\xi', \mbox{\boldmath 
$\beta'$}, \mbox{\boldmath $\gamma'$},0 
\right) & = & \prod^M_{k=1} \int D^2 \{ f\} M_1 \left( 
\theta^*,\xi^*,t;\theta',\xi',0;[f] \right) \nonumber \\ & & 
\times N_{1_{k}} \left( \beta^*_k, \gamma^*_k, t; \beta'_k, 
\gamma'_k, 0; [f^*] \right), 
\end{eqnarray}
where $M_1$ is the propagator for
\begin{equation}
H_{M_{1}} = {\hbar \omega \over 2} \left( \xi^* {\partial \over 
\partial \xi^*} - \theta^* {\partial \over \partial \theta^*} 
\right) + {i\hbar \omega \over 2} f(t) \left( \xi^* {\partial 
\over \partial \xi^*} - \theta^* {\partial \over \partial 
\theta^*} \right), 
\end{equation}
and $N_{1_{k}}$ is the propagator for
\begin{equation}
H_{N_{1k}} = \hbar \omega_k \left( \gamma^*_k {\partial \over 
\partial \gamma^*_k} - \beta^*_k {\partial \over \partial 
\beta^*_k} \right) + \hbar f^*(t) c_k \left( \gamma^*_k 
{\partial \over \partial \beta^*_k} + \beta^*_k {\partial \over 
\partial \gamma^*_k} \right). 
\end{equation}
Here the propagator $M_1$ is
\begin{equation}
M_1 = \sum\limits^{\infty}_{p=1} {1 \over p!} \left[ \left( 
\theta^*, \xi^* \right) \widetilde{Q} \pmatrix{\theta' \cr 
\xi'} \right]^p, 
\end{equation}
where $\widetilde{Q}$ is given by
\begin{equation}
\widetilde{Q} (t) = \exp \left( {i\omega \over 2} S_z t - 
{\omega \over 2} S_z \int\limits^t_0 d\tau f(\tau) \right) ,
\end{equation}
with $\widetilde{Q}_{ij}(0) = \delta_{ij}, \widetilde{Q}(0)=I$.

The propagator $N_{1_{k}}$ is
\begin{equation}
N_{1_{k}} = \sum\limits^{\infty}_{l=0} {1 \over l!} \left[ 
\left( \beta^*_k, \gamma^*_k \right) Q^k \pmatrix{\beta'_k \cr 
\gamma'_k} \right]^l, 
\end{equation}
where $Q^k$ satisfies the equation
\begin{equation}
{\partial \over \partial t} Q^k = i \left( \omega_k 
\sigma_{z_{k}} - f^* (t) c_k \sigma_{x_{k}} \right) Q^{(k)}. 
\end{equation}
This equation can be solved recursively to give 
\begin{equation}
Q^k (t) = \sum\limits^{\infty}_{n=0} Q^{k(n)} (t) ,
\end{equation}
with
\begin{equation}
Q^{k(0)} (0) = I,~~~ Q^{k(n)} (0) (n \neq 0) = 0,
\end{equation}
\begin{eqnarray}
Q^{k(n)} (t) & = & \left( i\omega_k \sigma_{z_{k}} \right)^n 
\int\limits^t_0 d\tau_n \int\limits^{\tau_{n}}_0 d\tau_{n-1} 
... \int\limits^{\tau_{2}}_0 d\tau_1 \nonumber \\ & & \times 
\exp \left( -i\sigma_{x_{k}} c_k \left( 
\int\limits^{\tau_{1}}_0 - \int\limits^{\tau_{2}}_{\tau_{1}} + 
... + (-1)^n \int\limits^t_{\tau_{n}} \right) d\tau f^* (\tau) 
\right). 
\end{eqnarray}
Using Eqs. (54), (56) with $p=1$, $l=1$, respectively, in 
Eq. (51) and making use of Eqs. (55), (58), (59), (60) along 
with the $\delta$-functional identity (13), the amplitudes of 
the propagator for the Hamiltonian (43) are obtained in matrix 
form (in the Hilbert space of $H_R$) as 
\begin{eqnarray}
\fl u_4 = \pmatrix{U_{00} & U_{01} \cr U_{10} & U_{11}} & = & 
\prod^M_{k=1} \sum\limits^{\infty}_{n=0} (i\omega_k)^n 
\int\limits^t_0 d\tau_n \int\limits^{\tau_{n}}_0 d\tau_{n-1}... 
\int\limits^{\tau_{2}}_0 d\tau_1 \nonumber\\ & & \times 
e^{i{\omega \over 2} S_z t} \pmatrix{\cos (\Theta^{k(n)}) & i 
\sin (\Theta^{k(n)}) \cr (-1)^ni \sin (\Theta^{k(n)}) & (-1)^n 
\cos (\Theta^{k(n)})}, 
\end{eqnarray}
where
\begin{equation}
\Theta^{k(n)} = {\omega \over 2} S_z c_k A_n,
\end{equation}
\begin{equation}
A_n = \sum\limits^n_{j=1} (-1)^{j+1} 2\tau_j + (-1)^nt.
\end{equation}

Now if we expand the terms containing $S_z$, i.e., make an 
expansion in the system space, in Eq. (61) we get terms such as 
\begin{equation}
e^{i{\omega \over 2}S_zt} \cos(\Theta^{k(n)}) = \cos({\omega 
\over 2} c_k A_n) \pmatrix{ e^{i{\omega \over 2}t} & o \cr 0 & 
e^{-i{\omega \over 2}t} }. 
\end{equation} 
Here we have used the fact that
\begin{equation}
e^{S_z A} = \pmatrix{ e^{A} & o \cr 0 & e^{-A} }.
\end{equation}
Similarly, 
\begin{equation}
e^{i{\omega \over 2}S_zt} i \sin(\Theta^{k(n)}) = i 
\sin({\omega \over 2} c_k A_n) \pmatrix{ e^{i{\omega \over 2}t} 
& o \cr 0 & -e^{-i{\omega \over 2}t} }. 
\end{equation} 
The above equations have only diagonal elements. We can see 
from the above equations that there are 16 amplitudes of the 
propagator for each mode $k$ of the reservoir out of which only 
the energy-conserving terms are present due to the QND nature 
of the system-reservoir coupling. 

\section{Discussions}

We look closely at the forms of the propagators (23) and (61) 
of the QND type Hamiltonians (\ref{h1}) and (43), respectively. 
In the first case with an oscillator bath, Eq. (23) involves 
the matrix 
\[ \pmatrix{e^B & 0 \cr 0 & e^{-B}}, \]
where $B$ is given by Eq. (25). This can be used to generate 
the following transformation in phase space: 
\begin{equation}
\pmatrix{X \cr P} = \pmatrix{e^B & 0 \cr 0 & e^{-B}} \pmatrix{x 
\cr p}. \label{s1} 
\end{equation}
It can be easily seen from Eq. (\ref{s1}) that the Jacobian of 
the transformation is unity and it is a phase space 
area-preserving transformation. The first matrix on the 
right-hand side of (\ref{s1}) has the form of a `squeezing' 
operation \cite{km91}, which is an area-preserving (in phase 
space) canonical transformation coming out as an artifact of 
homogeneous linear canonical transformations \cite{bk05}. 

In the second case of a spin bath, Eq. (61) involves the 
matrix 
\begin{equation}
R \equiv \pmatrix{\cos \Theta^{k(n)} & i \sin \Theta^{k(n)} \cr 
(-1)^ni \sin \Theta^{k(n)} & (-1)^n \cos \Theta^{k(n)}}, 
\label{h16} 
\end{equation}
where $\Theta^{k(n)}$ is given by Eq. (62). For 
particular $n$ and $k$, we write $\Theta^{k(n)}$ as $\Theta$. 
For $n$ even, the above matrix (\ref{h16}) becomes 
\begin{equation}
\pmatrix{\cos \Theta & i\sin \Theta \cr i\sin \Theta & \cos 
\Theta} = e^{i\Theta \sigma_x}.  \label{h17} 
\end{equation}
Using the Campbell-Baker-Hausdorff identity \cite{qsrl} this 
matrix can be shown to transform the spin vector $\sigma = 
(\sigma_x, \sigma_y, \sigma_z)$ as 
\begin{equation}
e^{i\Theta \sigma_x} \pmatrix{\sigma_x \cr \sigma_y \cr 
\sigma_z} e^{-i\Theta \sigma_x} = \pmatrix{1 & 0 & 0 \cr 0 & 
\cos 2\Theta & -\sin 2\Theta \cr 0 & \sin 2\Theta & \cos 
2\Theta} \pmatrix{\sigma_x \cr \sigma_y \cr \sigma_z},     
\label{h18} 
\end{equation}
i.e., the abstract spin vector is `rotated' about the $x$-axis 
by an angle $2\Theta$. For $n$ odd, (\ref{h16}) becomes (again 
writing $\Theta^{k(n)}$ for particular $n$ and $k$ as $\Theta$) 
\begin{equation}
\pmatrix{\cos \Theta & i\sin \Theta \cr -i\sin \Theta & -\cos 
\Theta} = \sigma_z \pmatrix{\cos \Theta & i\sin \Theta \cr 
i\sin \Theta & \cos \Theta} = \sigma_z e^{i\Theta \sigma_x}.   
\label{h19} 
\end{equation}
Thus the $n$-odd matrix is related to the $n$-even matrix by 
the spin-flipping energy. The above matrix transforms the spin 
vector $\sigma$ as 
\begin{equation}
\sigma_z e^{i\Theta \sigma_x} \pmatrix{\sigma_x \cr \sigma_y 
\cr \sigma_y} e^{-i\Theta \sigma_x} \sigma_z = e^{i\pi} 
\pmatrix{1 & 0 & 0 \cr 0 & \cos 2\Theta & \sin 2\Theta \cr 0 & 
\sin 2\Theta & -\cos 2\Theta} \pmatrix{\sigma_x \cr \sigma_y 
\cr \sigma_z}.  \label{r1} 
\end{equation}
It can be easily seen from the right-hand side of the 
Eq. (\ref{r1}) that the determinant of the transformation of 
the spin vectors brought about by the $n$-odd matrix 
(\ref{h19}) has the value unity. It is well known that the 
determinant of a rotation matrix is unity \cite{gms}. Thus we 
see that the above transformation has the form of a rotation. 
Specifically, it can be seen that 
\begin{equation}
\sigma_z \pmatrix{\cos 2\Theta & \sin 2\Theta \cr \sin 2\Theta 
& -\cos 2\Theta} = \pmatrix{ \cos 2\Theta & \sin 2\Theta \cr -
\sin 2\Theta & \cos 2\Theta}, 
\end{equation} 
and
\begin{equation}
\pmatrix{\cos 2\Theta & \sin 2\Theta \cr -\sin 2\Theta & \cos 
2\Theta}^T = \pmatrix{ \cos 2\Theta & -\sin 2\Theta \cr \sin 
2\Theta & \cos 2\Theta}.   \label{h20} 
\end{equation}
Here $T$ stands for the transpose operation. From the above it 
is seen that the matrix (\ref{h16}) has the form of the 
operation of `rotation', which is also a phase space 
area-preserving canonical transformation \cite{km91} and comes 
out as an artifact of homogeneous linear canonical 
transformations \cite{bk05}. Any element of the group of 
homogeneous linear canonical transformations can be written as 
a product of a unitary and a positive transformation 
\cite{barg, bk052}, which in turn can be shown to have unitary 
representations (in the Fock space) of rotation and squeezing 
operations, respectively \cite{bk05}. It is interesting that 
the propagators for the Hamiltonians given by Eqs. (\ref{h1}) 
and (43), one involving a two-level system coupled to a bath of 
harmonic oscillators and the other with a bath of two-level 
systems, are analogous to the squeezing and rotation 
operations, respectively. 

\section{Conclusions}

In this paper we have investigated the forms of the propagators 
of some QND Hamiltonians commonly used in the literature, for 
example, for the study of decoherence in quantum computers. We 
have evaluated the propagators using the functional integral 
treatment relying on coherent state path integration. We have 
treated the cases of a two-level system interacting with a 
bosonic bath of harmonic oscillators (section 2), and a spin 
bath of two-level systems (section 3). In each case the 
system-bath interaction is taken to be of the QND type, i.e., 
the Hamiltonian of the system commutes with the Hamiltonian 
describing the system-bath interaction. We have shown the 
commonly occuring free-particle coordinate coupling model to be 
unitarily equivalent to the free-particle velocity coupling 
model which is of the QND type. For the variants of the model 
in section 2, we have examined (a) the case where the 
two-level system in addition to interacting with the bosonic 
bath of harmonic oscillators is also acted upon by an external 
mode in resonance with the atomic transition (section 2.1), and 
(b) the non-QND spin-Bose problem (section 2.2), which could be 
used to describe the spin-Bose problem of the interaction of a 
two-level atom with the electromagnetic field modes in a cavity 
via a dipole interaction. 

The evaluation of the exact propagators of these many body 
systems could, apart from their technical relevance, also shed 
some light onto the structure of QND systems. We have found an 
interesting analogue of the propagators of these many-body 
Hamiltonians to squeezing and to rotation, for the bosonic and 
spin baths, respectively. Every homogeneous linear canonical 
transformation can be factored into the rotation and squeezing 
operations and these cannot in general be mapped from one to 
the other -- just as one cannot in general map a spin bath to 
an oscillator bath (or vice versa) -- but together they span 
the class of homogeneous linear canonical transformations and 
are `universal'. Squeezing and rotation, being artifacts of 
homogeneous linear canonical transformations, are both 
phase-space area-preserving transformations, and thus this 
implies a curious analogy between the energy-preserving QND 
Hamiltonians and the homogeneous linear canonical 
transformations.  This insight into the structure of the QND 
systems would hopefully lead to future studies into this 
domain. 

\ack

It is a pleasure to acknowledge useful discussions with Joachim 
Kupsch. The School of Physical Sciences, Jawaharlal Nehru 
University, is supported by the University Grants Commission, 
India, under a Departmental Research Support scheme.


\section*{References}
\begin{thebibliography}{99}


\bibitem{caves80} Caves C M, Thorne K D, Drever R W P, Sandberg 
V D and Zimmerman M 1980 {\it Rev. Mod. Phys.} {\bf 52} 341 

\bibitem{bo96} Bocko M F and Onofrio R 1996 {\it Rev. Mod. 
Phys.} {\bf 68} 755 

\bibitem{vo98} Onofrio R and Viola L 1998 {\it Phys. Rev. A} 
{\bf 58} 69 

\bibitem{zu84} Zurek W H 1984 {\it The Wave-Particle Dualism},  
eds Diner S, Fargue D, Lochak G and Selleri F (Dordrecht: D. 
Reidel Publishing Company) 

\bibitem{walls} Walls D F and Milburn G J 1994 {\it Quantum 
Optics} (Berlin: Springer); Milburn G J and Walls D F 1983 {\it 
Phys. Rev.} A {\bf 28} 2065 

\bibitem{science01} Orzel C, Tuchman A K, Fenselau M L, Yasuda 
M and Kasevich M A 2001 {\it Science} {\bf 291} 2386 

\bibitem{kbm98} Kuzmich A, Bigelow N P and Mandel L 1998 {\it 
Europhys. Lett.} {\bf 42} 481 

\bibitem{fkm65} Ford G W, Kac M and Mazur P 1965 {\it J. Math. 
Phys.} {\bf 6} 504 

\bibitem{cl83} Caldeira A O and Leggett A J 1983 {\it Physica 
A} {bf 121} 587 

\bibitem{wz91} Zurek W H 1991 {\it Phys. Today} {\bf 44} 36; 
Zurek W H 1993 {\it Prog. Theo. Phys.} {\bf 81} 28 

\bibitem{fv63} Feynman R P and Vernon F L 1963 {\it Ann. Phys. 
(N.Y.)} {\bf 24} 118 

\bibitem{ha85} Hakim V and Ambegaokar V 1985 {\it Phys. Rev.} A 
{\bf 32} 423 

\bibitem{sc87} Smith C M and Caldeira A O 1987 {\it Phys. Rev.} 
A {\bf 36} 3509; {\it ibid} 1990 {\bf 41} 3103 

\bibitem{gsi88} Grabert H, Schramm P and Ingold G L 1988 {\it 
Phys. Rep.} {\bf 168} 115 

\bibitem{sb00} Banerjee S and Ghosh R 2000 {\it Phys. Rev.} A 
{\bf 62} 042105 

\bibitem{sb03-2} Banerjee S and Ghosh R 2003 {\it Phys. Rev.} E 
{\bf 67} 056120 

\bibitem{gkd01} Gangopadhyay G, Kumar M S and Dattagupta S 2001 
{\it J. Phys. A: Math. Gen.} {\bf 34} 5485 

\bibitem{sgc96} Shao J, Ge M-L and Cheng H 1996 {\it Phys. 
Rev.} E {\bf 53} 1243 

\bibitem{rpp00} Prokof'ev  N V and Stamp P C E 2000 {\it Rep. 
Prog. Phys.} {\bf 63} 669 

\bibitem{schwin} Schwinger J 1965 {\it Quantum Theory of 
Angular Momentum} eds Biedenharn L C and von Dam H (New York: 
Academic) 

\bibitem{vb47} Bargmann V 1947 {\it Ann. Math.} {\bf 48} 568; 
Bargmann V 1962 {\it Rev. Mod. Phys.} {\bf 34} 829 

\bibitem{papa86} Papadapoulas G J 1986 {\it J. Math. Phys.} 
{\bf 27} 221; Erratum 1986 {\it J. Math. Phys.} {\bf 27} 1492; 
Papadapoulas G J 1985 {\it J. Phys. A: Math. Gen.} {\bf 18} 
1945 

\bibitem{lc87} Leggett A J, Chakravarty S, Dorsey A T, Fisher M 
P A, Garg A K and Zwerger W 1987 {\it Rev. Mod. Phys.} {\bf 59} 
1 

\bibitem{unruh95} Unruh W G 1995 {\it Phys. Rev.} A {\bf 51} 
992 

\bibitem{ps96} Palma G M, Suominen K-A and Ekert A K 1996 {\it 
Proc. R. Soc. Lond.} A {\bf 452} 567 

\bibitem{dd95} DiVincenzo D P 1995 {\it Phys. Rev.} A {\bf 51} 
1015 

\bibitem{sb03-1} Banerjee S and Ghosh R 2003 {\it J. Phys. A: 
Math. Gen.} {\bf 36} 5787 

\bibitem{klauder} Klauder J R 1960 {\it Ann. Phys. (N.Y.)} {\bf 
11} 123 

\bibitem{flc88} Ford G W, Lewis J T and O'Connell R F 1988 {\it 
Phys. Rev.} A {\bf 37} 4419 

\bibitem{cm78} Mavroyannis C 1978 {\it Phys. Rev.} A {\bf 18} 
185 

\bibitem{rn82} Reik H G, Nusser H and Amarante Ribeiro L A 1982 
{\it J. Phys. A} {\bf 15} 3491 

\bibitem{be70} Blume M, Emery V J and Luther A 1970 {\it Phys. 
Rev. Lett.} {\bf 25} 450 

\bibitem{cl84} Chakravarty S and Leggett A J 1984 {\it Phys. 
Rev. Lett.} {\bf 52} 5 

\bibitem{uw93} Weiss U 1999 {\it Quantum Dissipative Systems} 
(Singapore: World Scientific) 

\bibitem{sh97} Shao J and Hanggi P 1998 {\it Phys. Rev. Lett.} 
{\bf 81} 5710; and references therein. 

\bibitem{km91} Kim Y S and Noz M E 1991 {\it Phase Space 
Picture of Quantum Mechanics: Group Theoretical Approach} 
(Singapore: World Scientific). Note that $B$ appearing in 
(\ref{s1}) as defined in (25) is complex while in the usual 
discussions of squeezed states, it is real. 

\bibitem{bk05} Banerjee S and Kupsch J 2005 {\it J. Phys. A: 
Math. Gen.} {\bf 28} 5237 

\bibitem{qsrl} Louisell W H 1973 {\it Quantum Statistical 
Properties of Radiation} (New York: John Wiley and Sons) 

\bibitem{gms} Greiner W and Muller B 1989 {\it Quantum 
Mechanics: Symmetries} (Berlin: Springer-Verlag) 

\bibitem{barg} Bargmann V 1970 {\it Analytic Methods in 
Mathematical Physics} eds Gilbert R P and Newton R G (New York: 
Gordon and Breach) 

\bibitem{bk052} Kupsch J and Banerjee S 2006 
Infinite Dimensional Analysis, Quantum Probability and Related
Topics {\bf 9}, 413 (2006); Los Alamos Archive arXiv:math-ph/0410049.

\end{thebibliography}
\end{document}